%% file: main.tex
\documentclass{article}

\usepackage{microtype}
\usepackage{graphicx}
\usepackage{subcaption}
\usepackage{booktabs}
\usepackage{hyperref}

\usepackage[accepted]{icml2026}

\usepackage{amsmath,amssymb,mathtools}
\usepackage{bm}
\usepackage{multirow}
\usepackage{array}
\usepackage[capitalize,noabbrev]{cleveref}

\icmltitlerunning{Representation Matters in Randomized Smoothing for Audio Classification}

\newcommand{\va}{\bm{a}}

\newcommand{\vx}{\bm{x}}
\newcommand{\vz}{\bm{z}}

\newcommand{\vr}{\bm{r}}
\newcommand{\bI}{\bm{I}}
\newcommand{\veps}{\boldsymbol{\epsilon}}
\newcommand{\veta}{\boldsymbol{\eta}}
\newcommand{\vxi}{\boldsymbol{\xi}}
\newcommand{\vdelta}{\boldsymbol{\delta}}
\newcommand{\vDelta}{\boldsymbol{\Delta}}

\begin{document}
\twocolumn[
  \icmltitle{Representation Matters in Randomized Smoothing for Audio Classification}

  \begin{icmlauthorlist}
    \icmlauthor{Jong-Ik Park}{cmu}
    \icmlauthor{Shreyas Chaudhari}{cmu}
    \icmlauthor{Jos\'{e} M. F. Moura}{cmu}
    \icmlauthor{Carlee Joe-Wong}{cmu}
  \end{icmlauthorlist}

  \icmlaffiliation{cmu}{Carnegie Mellon University, Pittsburgh, Pennsylvania, USA}

  \icmlcorrespondingauthor{Jong-Ik Park}{jongikp@andrew.cmu.edu}

  \icmlkeywords{certified robustness, randomized smoothing, audio classification, audio robustness, representation ambiguity}
  \vskip 0.25in
]
\printAffiliationsAndNotice{}

\begin{abstract}
Randomized smoothing (RS) certifies robustness in the vector space where Gaussian noise is added. In audio classification, this space is often not uniquely defined as standard pipelines normalize, range-control, and transform waveforms into log-mel or other spectral features. We show that direct RS is therefore underspecified unless the certified object and preprocessing policy are explicit.
On two audio benchmarks, keyword spotting and environmental-sound classification, we study waveform, feature-space, and post-processed smoothing. Our diagnostics show why representation-aware reporting is necessary: at the same smoothing level $\sigma=0.0025$, the two datasets share the same median raw radius $.007996$, but different waveform energies yield different SNR-equivalent scales ($83.98$ vs. $90.97$ dB); log-mel smoothing gives higher positive-radius certified accuracy on environmental sounds ($68.42\%$ vs. $65.53\%$), certifying more examples with nonzero radius but over features rather than waveforms; and clipping or peak normalization changes the effective perturbation norm by roughly $230$--$351\times$.
We therefore recommend that audio RS studies choose and report the task-specific certified object and perturbation model, including the perturbation location, gain policy, raw radius, and any post-noise geometry changes.
\end{abstract}

\input{main_body/intro}
\input{main_body/related_work}
\input{main_body/motivation}

\input{main_body/methodology}

\input{main_body/experimental}

\section{Discussion and Conclusion}
\label{sec:discussion}

This paper shows that randomized smoothing for audio is not fully specified by a certified radius alone.
Since RS certifies the vector space where noise is added, waveform smoothing, feature-space smoothing, and post-processed smoothing define different robustness claims.
Our diagnostics show that these distinctions matter in practice: feature-space smoothing can look numerically favorable while certifying a different object, and post-noise preprocessing can change the perturbation geometry by orders of magnitude.
The main recommendation is therefore simple: audio RS studies should explicitly report the certified object, perturbation location, gain policy, raw radius, signal-relative scale, and any post-noise transformation.

\clearpage
\newpage

\section*{Acknowledgments}
This work was partially supported by the National Science Foundation (NSF) under Grant CCF-2327905.

\bibliography{mybib}
\bibliographystyle{icml2026}

\end{document}

%% file: main_body/intro.tex
\section{Introduction}
\label{sec:intro}

Audio models increasingly drive sensing pipelines---wake-word interfaces,
acoustic event detection, health monitoring, and industrial inspection of motors,
valves, pumps, and rails.
Industrial machine-listening datasets and challenges, such as Malfunctioning Industrial Machine Investigation and Inspection (MIMII)~\citep{purohit2019mimii} and Detection and Classification of Acoustic Scenes and Events (DCASE) anomalous-sound detection~\citep{dcase2025task2}, show the growing role of audio models in condition monitoring and factory automation.
In these settings, ordinary test accuracy is not enough: a model can be accurate on curated data but unstable under microphone gain changes, sensor drift, background noise, or malicious perturbations.

Randomized smoothing (RS) certifies robustness by classifying Gaussian-perturbed
inputs and certifying the majority-vote class within a provable radius
\citep{lecuyer2019certified,cohen2019certified,li2019certified}. Crucially, RS
certifies perturbations \emph{in the vector space where the noise is added}. For
images, this space is the normalized pixel tensor. For audio, it is far less
canonical: resample, window, gain-normalize, range-control, and
map waveforms to log-mel or other spectral features
\citep{piczak2015environmental,hershey2017cnn,kong2020panns,gong2021ast,
lostanlen2019pcen,zeghidour2021leaf}. Smoothing/normalizing a waveform, 
a log-mel tensor, or an embedding therefore yields guarantees that are each valid
but mutually incomparable.

We study this representation ambiguity. Direct RS is mathematically valid once the input vector is fixed, but as an audio reporting practice, it is underspecified unless the certified object and preprocessing policy are stated. Noise on a waveform certifies waveform perturbations; noise on a log-mel tensor certifies feature perturbations. And when a noisy waveform is clipped or renormalized, the resulting certificate is still valid, but it certifies a composed classifier rather than the original waveform classifier its radius may appear to describe. Two RS pipelines can thus report the same radius while certifying different objects.

Prior audio robustness work is largely empirical---noise, reverberation,
time-frequency augmentation, and adversarial attacks
\citep{snyder2015musan,ko2017reverberant,park2019specaugment,carlini2016hiddenvoice,
carlini2018audio,qin2019imperceptible,yuan2018commandersong}---and certified work
has extended RS to ASR and speaker recognition
\citep{olivier2021sequential,korzh2024speaker}, but none analyzes how routine
preprocessing changes the certified input space. We address this gap by studying
three representative contracts---waveform, feature-space, and post-processed
smoothing---and showing they yield valid but different certificates, necessitating a
representation-aware reporting framework.

\textbf{Contributions.}
(i) We identify an under-specification in direct RS reporting for audio: waveform
normalization, log-mel extraction, and post-noise range control each yield valid
but different certificates.
(ii) We formulate a representation-aware reporting framework specifying the
certified object, perturbation location, gain policy, raw radius, signal-relative
scale, and any post-noise transformation.
(iii) We give diagnostics showing how representation-agnostic reporting misleads:
identical raw radii can carry different signal-relative meaning; log-mel smoothing
can appear stronger while certifying features; and clipping or peak normalization
can change the effective perturbation geometry.

%% file: main_body/related_work.tex
\section{Background and Related Work}
\label{sec:background}

\textbf{Audio front ends.}
Audio classifiers process recordings through resampling, windowing, normalization, and spectral features rather than raw audio alone~\citep{warden2018speech,piczak2015esc}.
Large-scale audio event recognition and modern models further rely on learned representations, pretrained audio networks, and spectrogram transformers~\citep{gemmeke2017audioset,hershey2017cnn,piczak2015environmental,kong2020panns,gong2021ast}.
Learnable and normalized front ends, including Learnable Audio Frontend (LEAF) and per-channel energy normalization (PCEN), also show that representation design is part of the model pipeline~\citep{zeghidour2021leaf,lostanlen2019pcen}.
Thus, when applying RS to audio, the certified representation must be explicit.

\textbf{Empirical audio robustness.}
Audio robustness is commonly studied with augmentation or attack-based evaluation.
MUSAN noise, reverberation augmentation, and SpecAugment improve or evaluate robustness under sampled perturbations~\citep{snyder2015musan,ko2017reverberant,park2019specaugment}, while adversarial-audio work shows that speech systems can be manipulated by hidden or targeted commands~\citep{carlini2016hiddenvoice,carlini2018audio,qin2019imperceptible,yuan2018commandersong}.
These methods motivate robustness but do not certify prediction invariance under all perturbations inside a specified neighborhood.

\textbf{Randomized smoothing and certified audio.}
RS provides such a formal neighborhood guarantee by averaging predictions under Gaussian noise and certifying the top smoothed class~\citep{cohen2019certified}.
Prior work extends RS through differential privacy, additive noise, adversarial training, $f$-divergence, and projected certificates~\citep{lecuyer2019certified,li2019certified,salman2019provably,dvijotham2020framework,pfrommer2023projected}.
Certified audio work has adapted RS to ASR sequence outputs and speaker recognition~\citep{olivier2021sequential,korzh2024speaker}.
We study a complementary issue: even when RS applies directly, audio preprocessing can change what input space is being certified. Our focus is therefore not a new smoothing theorem, but a representation-aware reporting framework for audio RS.

%% file: main_body/motivation.tex
\section{Why Direct RS Is Under-Specified for Audio}
\label{sec:motivation}

Randomized smoothing certifies perturbations in the space where noise is added.
For a classifier $f$ and noise $\veps\sim\mathcal{N}(\bm{0},\sigma^2\bI)$, the smoothed classifier is
\begin{equation}
    g(\vx)=\arg\max_k \Pr_{\veps}\!\big[f(\vx+\veps)=k\big].
    \label{eq:smoothed_classifier}
\end{equation}
Given a lower confidence bound $\underline{p}_A>1/2$ on the top class probability, $g$ is constant on the $\ell_2$ ball of radius
\begin{equation}
    R=\sigma\Phi^{-1}(\underline{p}_A)\quad~\text{\cite{cohen2019certified}},
    \label{eq:rs_radius}
\end{equation}
where $\Phi^{-1}$ is the inverse standard Gaussian CDF.
Thus, the radius is interpretable only where noise is added.

For audio, that space is often ambiguous.
A raw recording may be resampled, gain-normalized, cropped or padded, range-controlled, and mapped to spectral features before classification.
Let $\vx$ be the waveform, $\vz=\psi(\vx)$ its log-mel feature, $h$ a feature classifier, and $\mathcal{T}$ a post-noise transformation such as clipping or normalization.
Three common direct RS choices are
\begin{equation}
    \underbrace{f(\vx+\veps)}_{\text{waveform}},
    \qquad
    \underbrace{h(\vz+\veta)}_{\text{feature}},
    \qquad
    \underbrace{f(\mathcal{T}(\vx+\veps))}_{\text{post-processed}},
    \label{eq:three_smoothings}
\end{equation}
where $\veps$ is waveform-space noise and $\veta$ is feature-space noise.
These choices are all valid RS applications, but they certify different objects.

\textbf{Failure mode 1: feature certificates are not waveform certificates.}
Waveform RS certifies a waveform ball, and Feature RS instead certifies a feature-space ball,
\begin{equation}
\begin{aligned}
    &\mathcal{B}_{\rm wav}(\vx,R)
    =
    \{\vx+\vdelta:\|\vdelta\|_2\le R\}. \\
    &\mathcal{B}_{\rm feat}(\vz,R_z)
    =
    \{\vz+\vxi:\|\vxi\|_2\le R_z\}.
    \label{eq:wav_ball}
\end{aligned}
\end{equation}
Since log-mel features are nonlinear and non-invertible, this feature ball does not directly define a waveform ball:
\begin{equation}
    \psi^{-1}\!\left(\mathcal{B}_{\rm feat}(\psi(\vx),R_z)\right)
    \neq
    \mathcal{B}_{\rm wav}(\vx,R)
    \label{eq:feature_not_waveform}
\end{equation}
for any directly comparable $R$.
Thus, a feature certificate may have a larger numerical certified accuracy or radius while certifying a different object.

\textbf{Failure mode 2: raw radius omits signal scale.}
A raw radius is an absolute coordinate-space quantity, whereas audio perturbation scale is usually interpreted relative to the signal. Consider scaling the waveform by a gain $c>0$, i.e.\ $\vx\mapsto c\vx$. The norm scales accordingly, $\|c\vx\|_2=c\|\vx\|_2$, so the signal-relative meaning of a fixed raw radius $R$,
\begin{equation}
    \mathrm{SNR}_{\rm cert}(\vx,R)=20\log_{10}\frac{\|\vx\|_2}{R},
    \label{eq:snr_definition_motivation}
\end{equation}
shifts under the gain change by $c$ as
$
    \mathrm{SNR}_{\rm cert}(c\vx,R)=\mathrm{SNR}_{\rm cert}(\vx,R)+20\log_{10}c.
$
Common normalization can also remove gain before certification; for RMS normalization, $\mathcal{R}_{\rm rms}(c\vx)=\mathcal{R}_{\rm rms}(\vx)$, for $ c>0$. Thus, a radius reported without the gain policy and signal-relative scale is not audio-interpretable.

\textbf{Failure mode 3: post-noise preprocessing certifies a composed classifier.} 
Direct waveform RS nominally evaluates $\vx+\veps$, but if the implementation clips or renormalizes the noisy waveform, the model instead receives 
\begin{equation}
    \mathcal{T}(\vx+\veps)=\vx+\vDelta_{\mathcal{T}},
    \quad
    \vDelta_{\mathcal{T}}=\mathcal{T}(\vx+\veps)-\vx,
    \label{eq:effective_delta_motivation}
\end{equation}
where in general $\vDelta_{\mathcal{T}}\neq\veps$ and is non-Gaussian. RS remains valid here as it instead smooths the composed classifier $f\circ\mathcal{T}$. \Cref{eq:rs_radius} still certifies constancy on the additive ball $\mathcal{B}_{\rm wav}(\vx,R)$ in waveform space. The certified region is thus the same $\ell_2$ ball as in \Cref{eq:wav_ball} and there is no guarantee over a transformed set $\{\mathcal{T}(\vx+\vdelta):\|\vdelta\|_2\le R\}$, which need not even be a ball. What changes is the certified function, not the certified set. A radius presented as a waveform certificate appears to bound $f$ under additive perturbations of $\vx$, but in fact bounds $f\circ\mathcal{T}$. The two agree only when $\vDelta_{\mathcal{T}}=\veps$; otherwise the certificate concerns a different map, a discrepancy we quantify with $D_{\rm geom}$.

Therefore, the limitation of direct reporting is not that the RS theorem is invalid, but a bare radius or certified accuracy does not say whether the certificate concerns waveform perturbations, feature perturbations, or a post-processed composed classifier.
Audio RS results should therefore report the certified object, perturbation location, gain policy, signal-relative scale, and any post-noise transformation.

%% file: main_body/methodology.tex
\section{Representation-Aware Audio RS}
\label{sec:method}

This section converts the failure modes in \Cref{sec:motivation} into a reporting contract that specifies the certified object, perturbation location, gain policy, and any post-noise preprocessing.

\textbf{Reporting contract.}
An audio RS certificate should specify 
$
(\rho,\;\mathcal{P},\;\mathcal{T},\;m),
$
where $\rho$ maps raw audio $\va$ to the certified vector $\vr=\rho(\va)$,
$\mathcal{P}$ is the perturbation applied to $\vr$, $\mathcal{T}$ any
post-perturbation transformation, and $m$ the reporting scale (raw radius or
SNR-equivalent). Since RS certifies an $\ell_2$ ball in the space of $\vr$, the
representation should make that ball match the perturbation the application must
tolerate. 

Sensor- and channel-level threats, such as gain, acoustic noise, and transmission corruption, are additive waveform perturbations, calling for waveform $\rho$ with identity $\mathcal{T}$ (Contract~1), reported on the SNR-equivalent scale. Feature front-end stability is certified in feature space (Contract~2), but reported as feature, not waveform, robustness. A pipeline that clips or renormalizes at inference certifies $f\circ\mathcal{T}$ (Contract~3), which must be labeled as such since it differs from $f$ on additive inputs. In short, we recommend fixing the perturbation model first, choosing $(\rho,\mathcal{P},\mathcal{T})$ to express it.

\textbf{Contract 1: fixed-gain waveform smoothing.}
For sensor-level perturbations, we use a fixed-gain waveform vector
\begin{equation}
    \vx=s(\va)\,\mathrm{CropPad}(\mathrm{Resample}(\va)),
    \label{eq:fixed_wave_x}
\end{equation}
where $\mathrm{Resample}$ converts raw audio to the target sampling rate, $\mathrm{CropPad}$ deterministically crops or zero-pads the signal to length $T$, and $s(\va)$ is a scalar gain computed once from the clean segment.
This contract uses $\rho(\va)=\vx$, waveform Gaussian noise
$\mathcal{P}=\mathcal{N}(\bm{0},\sigma^2\bI_T)$, and identity post-processing $\mathcal{T}_{\rm fixed}(\tilde{\vx})=\tilde{\vx}$.
The radius is the standard RS radius in \Cref{eq:rs_radius}, computed with a one-sided Clopper--Pearson lower bound~\citep{clopper1934use}.
When the lower bound is above $1/2$, the certificate is over the waveform ball in \Cref{eq:wav_ball}.
Because this radius is an absolute waveform-space distance, we report the SNR-equivalent scale defined in \Cref{eq:snr_definition_motivation}.

\textbf{Contract 2: feature-space smoothing.}
Feature smoothing uses $\rho(\va)=\vz=\psi(\vx)$ and adds Gaussian noise in feature space.
It therefore certifies the feature-space ball, not the waveform ball in \Cref{eq:wav_ball}.
This contract can be useful for feature stability, but by \Cref{eq:feature_not_waveform} it should not be reported as waveform robustness.

\textbf{Contract 3: post-processed waveform smoothing.}
Post-processed smoothing uses $\rho(\va)=\vx$, adds waveform Gaussian noise, and then applies a post-noise transformation $\mathcal{T}$ before classification.
In our diagnostics, $\mathcal{T}$ is one of four policies: \\
\indent$
    \quad\mathcal{T}_{\rm fixed}(\tilde{\vx})
    = \tilde{\vx}, 
    \qquad\quad\qquad\;\;
    \mathcal{T}_{\rm rms}(\tilde{\vx})
    = \tilde{\vx}\frac{\mathrm{RMS}(\vx)}{\mathrm{RMS}(\tilde{\vx})},$ \\
\indent$
    \quad\mathcal{T}_{\rm clip}(\tilde{\vx})
    = \mathrm{clip}(\tilde{\vx},-1,1), 
    \quad
    \mathcal{T}_{\rm peak}(\tilde{\vx})
    = \tilde{\vx}\frac{\|\vx\|_{\infty}}{\|\tilde{\vx}\|_{\infty}},
$
\\
where $\tilde{\vx}=\vx+\veps$.
Clipping is applied sample-wise
$
    \mathrm{clip}(\tilde{x}_t,-1,1)
    =
    \min\{\max\{\tilde{x}_t,-1\},1\}.
$
The effective perturbation is $\vDelta_{\mathcal{T}}$ from
\Cref{eq:effective_delta_motivation}; as established in \Cref{sec:motivation}, RS
remains valid but certifies the composed classifier $f\circ\mathcal{T}$ on the
waveform ball of \Cref{eq:wav_ball}, rather than $f$ on additive perturbations.

To quantify how far $\mathcal{T}$ moves the effective perturbation from the
intended additive noise, we report
\begin{equation}
    D_{\rm geom}
    =
    \mathbb{E}_{\vx,\veps}\!\left[
    \frac{\big|\,\|\vDelta_{\mathcal{T}}\|_2-\|\veps\|_2\,\big|}{\|\veps\|_2}
    \right].
    \label{eq:dgeom}
\end{equation}
$D_{\rm geom}$ is not a certificate but an implementation diagnostic.
$D_{\rm geom}=0$ exactly when $\mathcal{T}$ leaves the perturbation unchanged
($\vDelta_{\mathcal{T}}=\veps$), so $f\circ\mathcal{T}$ matches the additive
classifier. Any nonzero value shows the two differ, with larger values measuring
how far $f\circ\mathcal{T}$ departs from the additive-waveform classifier its
radius appears to describe.

%% file: main_body/experimental.tex
\section{Diagnostic Experiments}
\label{sec:experiments}

\textbf{Goal.}
The experiments test what a direct RS-style report would miss.
A direct report gives certified accuracy and radius, but may not specify the certified object, perturbation location, gain policy, or post-noise transformation.
We therefore evaluate the three contracts in \Cref{sec:method}: fixed-gain waveform smoothing, log-mel feature smoothing, and post-processed waveform smoothing.

\textbf{Setup.}
We evaluate keyword spotting on Speech Commands and environmental-sound classification on ESC-50.
Speech Commands uses 12 classes with $5{,}074$ test clips over three seeds; ESC-50 follows the official five-fold protocol with $400$ test clips per fold, averaged over three seeds and five folds. Both datasets consist of real recorded audio. All audio is mono, resampled to $16$ kHz, RMS-normalized, and cropped or padded to one second for Speech Commands and five seconds for ESC-50.
The classifier is a waveform-input log-mel CNN with $64$ mel bins, FFT size $1024$, window length $400$, hop length $160$, and four convolutional blocks with channels $32$--$64$--$128$--$128$.
Models use AdamW, learning rate $10^{-3}$, weight decay $10^{-4}$, gradient clipping $1.0$, cosine decay, bfloat16 mixed precision, and Gaussian waveform augmentation with standard deviation $0.005$.
Speech Commands is trained for $30$ epochs and ESC-50 for $200$.

\textbf{Certification.}
We use fixed-budget Monte Carlo RS with $n_0=100$ class-selection samples, $n=10{,}000$ certification samples, failure level $\alpha=0.001$, and $\sigma\in\{0.0025,0.005,0.01,0.02\}$.
Positive-radius certified accuracy is the fraction of test examples correctly classified by the smoothed classifier with nonzero certified radius.

\textbf{Reference: fixed-gain waveform certificates.}
\Cref{tab:main} reports waveform-space certificates, which serve as the reference for the diagnostics.
On keyword spotting, positive-radius certified accuracy stays between $92.10\%$ and $92.65\%$, while median certified SNR decreases from $83.98$ dB to $65.92$ dB as $\sigma$ increases.
On environmental sounds, certified accuracy decreases from $68.70\%$ to $65.53\%$, while median SNR decreases from $90.97$ dB to $72.91$ dB.
These results show that fixed-gain waveform smoothing gives nontrivial certificates; the main diagnostic question is whether direct reports remain meaningful without representation context. 

\begin{table}[t]
\centering
\caption{Fixed-gain waveform certificates. Positive-radius accuracy is the fraction correct with nonzero radius. Lower median SNR is stronger.}
\label{tab:main}
\resizebox{0.6\columnwidth}{!}{%
\begin{tabular}{lccc}
\toprule
Dataset, $\sigma$ & Pos.-rad. acc. & Med. $R$ & Med. SNR \\
\midrule
SC, .0025 & $92.25{\pm}0.57$ & .007996 & 83.98 \\
SC, .005  & $92.55{\pm}0.57$ & .015993 & 77.96 \\
SC, .01   & $92.65{\pm}0.61$ & .031986 & 71.94 \\
SC, .02   & $92.10{\pm}0.49$ & .063972 & 65.92 \\
\midrule
ESC, .0025 & $68.70{\pm}1.83$ & .007996 & 90.97 \\
ESC, .005  & $68.53{\pm}2.46$ & .015993 & 84.95 \\
ESC, .01   & $68.17{\pm}3.02$ & .031986 & 78.93 \\
ESC, .02   & $65.53{\pm}2.95$ & .063972 & 72.91 \\
\bottomrule
\end{tabular}}
\end{table}

\textbf{Failure mode 1: direct certified-accuracy comparison can choose the wrong object.}
\Cref{tab:diag} compares waveform smoothing with log-mel smoothing.
On ESC-50 at $\sigma=0.02$, a direct RS-style report would make log-mel smoothing appear preferable because it gives $68.42\%$ positive-radius certified accuracy, compared with $65.53\%$ for waveform smoothing.
This conclusion is incomplete: the log-mel certificate is over the feature tensor $\vz$, not waveform perturbations of $\vx$.
On Speech Commands, log-mel smoothing is lower ($91.59\%$ vs. $92.10\%$), confirming that the issue is not which representation always gives the larger number.
The issue is that the certified object changes.

\textbf{Failure mode 2: direct preprocessing can certify a transformed perturbation set.}
At $\sigma=0.0025$, clipping after noise reduces positive-radius certified accuracy from $92.25\%$ to $80.05\%$ on Speech Commands and from $68.70\%$ to $53.70\%$ on ESC-50.
A direct report might treat this as just another certified-accuracy number.
However, \Cref{tab:geom} shows that clipping and peak normalization change the effective perturbation norm by factors in the hundreds, while fixed-gain waveform smoothing preserves the intended Gaussian perturbation with effective norm ratio $1.0$.
Thus, these post-noise policies are not merely implementation details; they change the perturbation set being certified.

\begin{table}[t]
\centering
\caption{Representation and preprocessing diagnostics. Values are positive-radius certified accuracy percentages.}
\label{tab:diag}
\resizebox{\columnwidth}{!}{%
\begin{tabular}{lccc}
\toprule
Diagnostic & SC & ESC-50 & Direct-report failure \\
\midrule
Log-mel vs. wave, $\sigma=.02$ & $91.59$ vs $92.10$ & $68.42$ vs $65.53$ & certified object differs \\
Clip vs. wave, $\sigma=.0025$ & $80.05$ vs $92.25$ & $53.70$ vs $68.70$ & perturbation set differs \\
\bottomrule
\end{tabular}}
\end{table}

\begin{table}[t]
\centering
\caption{Perturbation geometry at $\sigma=0.0025$. Effective norm ratio is $\|\vDelta\|_2/\|\veps\|_2$.}
\label{tab:geom}
\resizebox{0.6\columnwidth}{!}{%
\begin{tabular}{lccc}
\toprule
Dataset & Protocol & $D_{\rm geom}$ & Eff. ratio \\
\midrule
SC & fixed gain & $6.49{\times}10^{-8}$ & 1.000 \\
SC & RMS renorm. & $3.38{\times}10^{-5}$ & 0.999985 \\
SC & clipping & 264.27 & 271.65 \\
SC & peak renorm. & 346.01 & 351.03 \\
\midrule
ESC & fixed gain & $3.28{\times}10^{-8}$ & 1.000 \\
ESC & RMS renorm. & $8.75{\times}10^{-6}$ & 0.999996 \\
ESC & clipping & 238.81 & 230.64 \\
ESC & peak renorm. & 346.80 & 350.61 \\
\bottomrule
\end{tabular}}
\end{table}